\documentclass[conference]{IEEEtran}

\usepackage{fancyhdr}
\usepackage{cite}
\usepackage{amsmath,amssymb,amsfonts}
\usepackage{algorithmic}
\usepackage{graphicx}
\usepackage{textcomp}
\usepackage[dvipsnames]{xcolor}
\usepackage[hyphens]{url}
\usepackage{multirow}
\usepackage{tabularx}
\usepackage{xspace}
\usepackage{enumitem}
\usepackage{siunitx}

\def\BibTeX{{\rm B\kern-.05em{\sc i\kern-.025em b}\kern-.08em
    T\kern-.1667em\lower.7ex\hbox{E}\kern-.125emX}}

\newif\ifanonymous
\anonymousfalse  
\newcommand{\anon}[2]{\ifanonymous #2\else #1\fi}
\newcommand{\redact}[1]{\anon{#1}{[redacted]}}
\newcommand{\anoncite}[1]{\ifanonymous\cite{redacted}\else\cite{#1}\fi}

\iffalse

\else

\fi

\ifanonymous
  \newcommand{\athena}{[redacted]\xspace}
\else
  \newcommand{\athena}{MTIA 300\xspace}
\fi

\newcommand{\hccl}{HCCL\xspace}
    
\begin{document}

\bstctlcite{IEEE:BSTcontrol}

\title{HCCL: Collective Communication for \redact{Meta Training and Inference Accelerators}}


\ifanonymous
\author{Anonymous Submission}
\else
\author{
\IEEEauthorblockN{
Wesley Bland\textsuperscript{1†},
Tiago Antunes\textsuperscript{2*},
Lars Paul Huse\textsuperscript{2†},
Chidambaram Muthu\textsuperscript{4†}, 
Adel Abouchaev\textsuperscript{4*}, \\
Rabib Alam\textsuperscript{1},
Abdullah Alperen\textsuperscript{1},
Alexey Andronov\textsuperscript{2*}, 
Jose Anto Akkara\textsuperscript{2*},
Vineet Badhwar\textsuperscript{4}, \\
Pavan Balaji\textsuperscript{1}, 
Daniel Berkovitch\textsuperscript{3*},
Bartosz Bogdanski\textsuperscript{2}, 
Shmeelok Chakraborty\textsuperscript{1}, 
Sungjun Cho\textsuperscript{1},
John Choi\textsuperscript{1}, \\
James Custer\textsuperscript{1}, 
Rodrigo De Castro\textsuperscript{4},
Nguyen Dinh Pham\textsuperscript{1*},
Matthew Edwards\textsuperscript{3}, 
Kristian Evensen\textsuperscript{2}, 
Evan Ezell\textsuperscript{1}, \\
Alex Finestead\textsuperscript{3}, 
Seth Goldstein\textsuperscript{3},
Prankur Gupta\textsuperscript{1}, 
Ranwei Hu\textsuperscript{3},
Adam Incera\textsuperscript{3*}, 
Anand Jayaraman\textsuperscript{1}, \\
Prashanth Kannan\textsuperscript{1}, 
Soumil Kanwal\textsuperscript{2}, 
Martin Karp\textsuperscript{2}, 
Sameer Kumar\textsuperscript{1}, 
Naina Kuruballi Mahesh\textsuperscript{5},
Wei Lin Guay\textsuperscript{2}, \\
Cristian Lumezanu\textsuperscript{3}, 
Cory Modlin\textsuperscript{1*}, 
Dag Georg Moxnes\textsuperscript{2}, 
Hoang Nam Nguyen\textsuperscript{1*}, 
Ashay Narsale\textsuperscript{1}, \\
Jaden Padua\textsuperscript{1*}, 
Kirtesh Patil\textsuperscript{1}, 
Minh Pham\textsuperscript{1},
Amin Qassoud\textsuperscript{1}, 
Ashwin Ramachandran\textsuperscript{8}, 
David Ramon Prados\textsuperscript{7*}, \\
Pallavi Shurpali\textsuperscript{1*},
Gregory R. Steinbrecher\textsuperscript{1*}, 
John Sundharam\textsuperscript{6}, 
Vangelis Tasoulas\textsuperscript{2}, 
Fuhou Tian\textsuperscript{4}, \\
Srinivas Vaidyanathan\textsuperscript{1}, 
Vimal Vasudevan\textsuperscript{1}, 
Nicolaas Viljoen\textsuperscript{1*}, 
Daniel Winkelman\textsuperscript{1}, 
Yijing Zeng\textsuperscript{4}, 
Zhaoqi Zhu\textsuperscript{1}, \\
Stig Arne Olsen\textsuperscript{2},
Gilad Goldfarb\textsuperscript{1*}, 
Rajiv Krishnamurthy\textsuperscript{1}, 
Rajeev Nair\textsuperscript{1}, 
Jonas Olsson\textsuperscript{2}, 
Joseph Provine\textsuperscript{1}, \\
Sreeram Ravinoothala\textsuperscript{1}, 
Shivayogi Ugaji\textsuperscript{1}, 
Hongyi Zeng\textsuperscript{1},
Nairan Zhang\textsuperscript{1}
}

\IEEEauthorblockA{
\textit{Meta Platforms}\\
\textsuperscript{†}Corresponding authors,
\textsuperscript{*}Work performed while employed at Meta Platforms \\
\textsuperscript{1}Menlo Park, CA, USA, 
\textsuperscript{2}Oslo, Norway,
\textsuperscript{3}New York, NY, USA,
\textsuperscript{4}Bellevue, WA, USA, 
\textsuperscript{5}Vancouver, BC, Canada, \\
\textsuperscript{6}London, UK,
\textsuperscript{7}Toronto, ON, Canada,
\textsuperscript{8}Austin, TX, USA\\
\{wbland, 
larsph, 
cmuthu\}@meta.com
}
}
\fi

\maketitle

\thispagestyle{fancy}
\lhead{}
\rhead{}
\chead{}
\cfoot{}
\renewcommand{\headrulewidth}{0pt}
\renewcommand{\footrulewidth}{0pt}


\begin{abstract}
We present HCCL, a collective communication library co-designed with \redact{Meta}'s \athena accelerator, the first \redact{Meta} chip to integrate backend networking directly on the chip package. \athena includes dedicated message engines (MEs) with near-memory compute (NMC) that fully offload collective execution from the compute grid, enabling large overlap between computation and communication. HCCL uses a compiled communication model in which the host generates a complete description of each collective including dependencies. We describe the control and data path architecture, topology-aware algorithm selection across \athena's asymmetric scale-up and scale-out network, and optimizations for both training and inference workloads. For training, HCCL achieves up to 940~GB/s on intra-rack collectives while introducing less than 0.5\% degradation to concurrent compute throughput. For inference, we leverage one-sided communication primitives that bypass the scheduling path to minimize collective latency and describe collective designs that improve compute-communication pipelining for latency-sensitive workloads.
\end{abstract}
\section{Introduction}
\label{sec:intro}

Ranking and recommendation models serve billions of users across several products~\cite{ne, dlrm, dhen, hstu}. 
These models are characterized by large embedding tables, irregular communication patterns dominated in time by AllToAllv, and a need to scale training across hundreds of accelerators. 
As these models have grown in complexity, communication has become an increasing fraction of overall training time, making the efficiency of the collective communication library a first-order concern for end-to-end performance.

To address this, \redact{Meta} developed the \redact{MTIA} chip family~\anoncite{freya, artemis, mtia-blog}, a line of custom accelerators designed specifically for the compute, memory, and communication demands of ranking and recommendation workloads.
\athena is the third generation in this line and the first to integrate backend networking directly on the chip package~\anoncite{athena_isca}. 
This integration eliminates the host-device-NIC communication present in traditional GPU-NIC architectures and enables a different approach to collective communication: one where the entire collective operation is offloaded to dedicated hardware on the accelerator itself.

This paper describes the Hoot Collective Communication Library (HCCL) built for \athena.
HCCL is co-designed with the chip's hardware capabilities, in particular its message engines (MEs), near-memory compute (NMC) units, and custom RDMA NIC IP blocks.
Rather than executing collectives as device compute kernels---the approach taken by libraries such as NCCL~\cite{nccl} and RCCL~\cite{rccl}---HCCL uses a compiled communication model in which the host generates a complete description of the collective as a set of subgraphs.
These subgraphs are then dispatched to the MEs for autonomous execution, freeing the compute grid entirely for application work.

We elaborate on the hardware communication components and how the communications software stack was co-designed to expose the best performance and functionality, focusing on the impact on application needs for both training models and more latency-sensitive workloads such as inference.

The key contributions of this paper are:

\begin{itemize}[leftmargin=*]
    \item A description of \athena's communication hardware, including the message engine architecture, near-memory compute for line-rate reductions, express doorbells for low-latency WQE submission, and the asymmetric scale-up/scale-out network design (Section~\ref{sec:hardware_architecture}).

    \item The design of HCCL's control and data path, including its compiled subgraph model, dynamic queue pair (QP) management under express doorbell constraints, and integration with PyTorch through both c10d and torchcomms backends (Section~\ref{sec:comms_architecture}).

    \item Optimizations for both training and inference workloads, including parallel ME processing for bandwidth-bound collectives, one-sided communication primitives for latency-sensitive inference, and device-resident and device-triggered collective designs that improve compute-communication pipelining (Section~\ref{sec:optimizations}).

    \item A performance evaluation demonstrating up to 940~GB/s collective bandwidth within a rack, near-zero compute degradation during overlapped execution, and sub-6~$\mu$s collective latency for inference-optimized paths (Section~\ref{sec:performance}).
\end{itemize}

\section{Hardware Architecture}
\label{sec:hardware_architecture}


\athena is \redact{Meta}'s first chip with onboard backend networking, requiring additional effort to efficiently enable these communication capabilities.
In this section, we describe both the hardware and software capabilities of \athena and how they were co-designed to provide a holistic solution for future training and inference workloads.


\subsection{\athena and System Architecture}


\begin{figure}[t]
    \centering
    \includegraphics[width=1\linewidth]{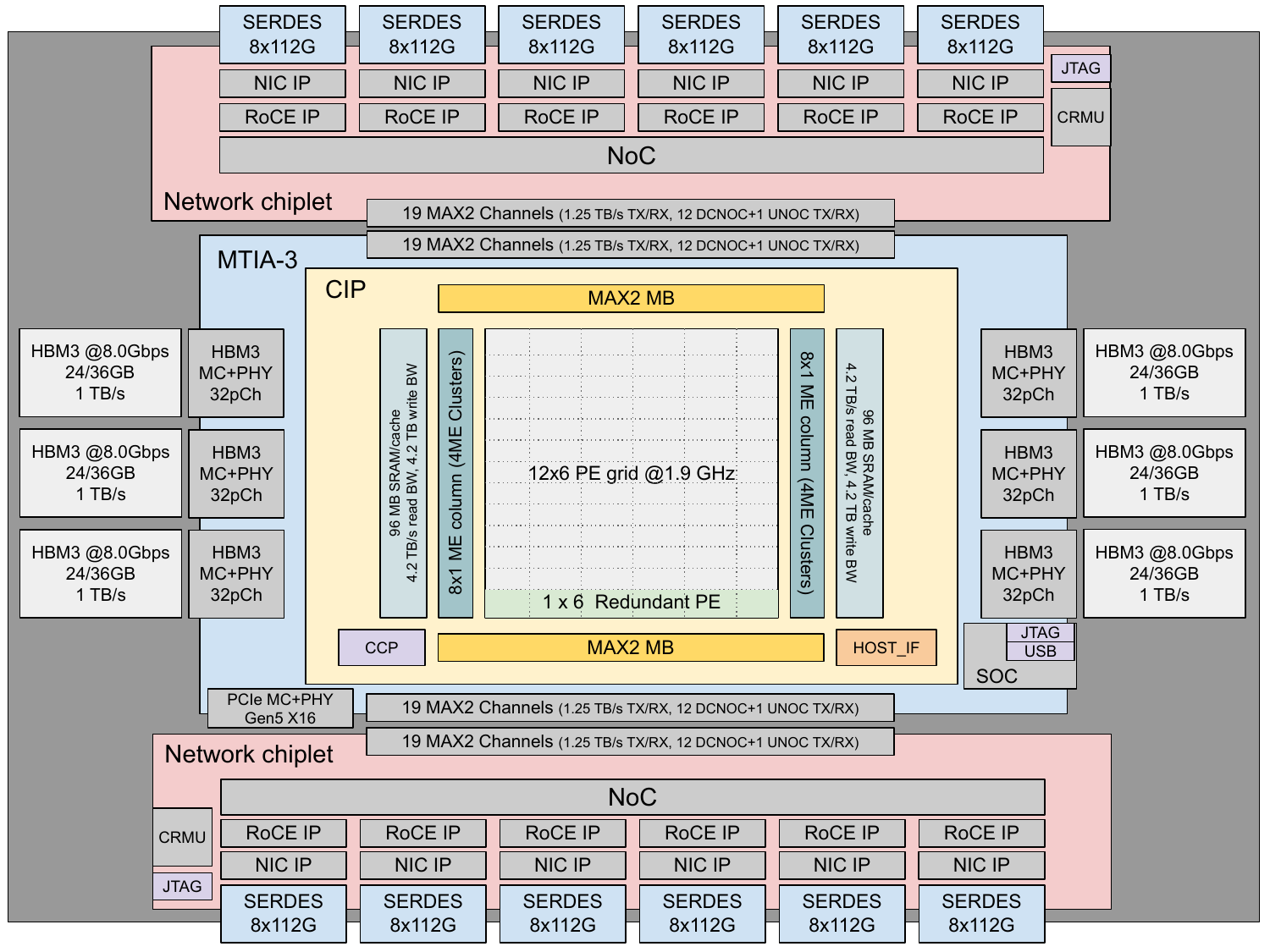}
    \caption{The \athena package with compute, network, and HBM.}
    \label{fig:chip-architecture}
\end{figure}

Figure~\ref{fig:chip-architecture} shows a high-level view of the entire \athena chip and its chiplet architecture.
Since it is optimized for recommendation models, \athena emphasizes High Bandwidth Memory (HBM) and networking over FLOPS.
A full analysis of the hardware design and its tradeoffs can be found in~\anoncite{athena_isca} where there is much more detail about the compute capabilities and comparisons with previous \redact{MTIA} chips.
In this paper, we focus primarily on the I/O capabilities and the network design and how it impacts collective communication.

\begin{figure}[t]
    \centering
    \includegraphics[width=1\linewidth]{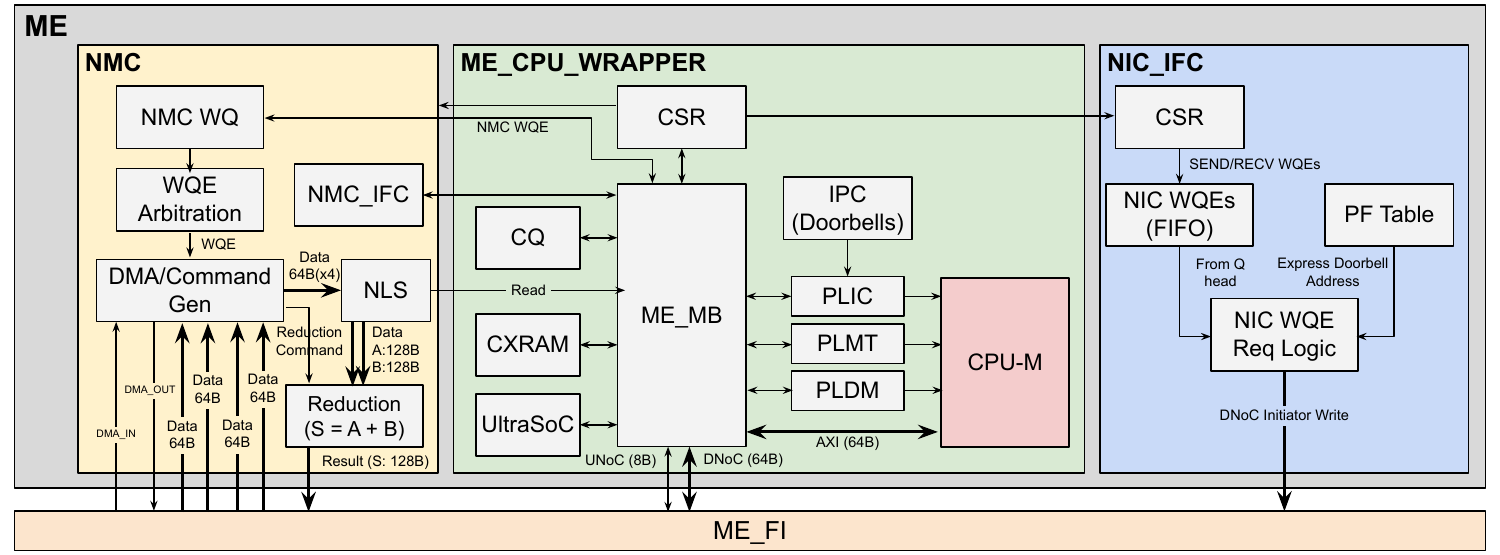}
    \caption{Message Engine (ME) architecture.}
    \label{fig:me-arch}
\end{figure}


\subsubsection{Control Core (CPU-C)}

Before any work is run on the compute or communication portion of the \athena chip, it is first submitted through a stream to the CPU-C to ensure it is ready to be dispatched and that the place where it will be executed is ready to receive it.
The CPU-C (shown in purple in Figure~\ref{fig:chip-architecture} inside the CCP module) is responsible for maintaining ordering within the stream and via dependencies expressed across streams.
Each of these streams can hold a mix of both compute and communication operations.
For communication, this means the CPU-C checks the description of each work packet and schedules each subgraph on the requested ME (see Section~\ref{subsec:hccl} for more details about collective construction).

\subsubsection{Message Engine (ME)}\label{subsubsec:me}
The \athena chip contains two groups of eight message engines which are used to manage the chip's network communications and collective compute. MEs bridge the gap between the subgraph representation of collectives and the representation used by the NICs with RDMA queue pairs and work requests. Each ME contains three distinct areas to accomplish different functions, the details of which can be seen in Figure~\ref{fig:me-arch} and are discussed below.

\paragraph{CPU-M} The first component (in red with related components in green) is a RISC-V communication core (CPU-M) that handles subgraphs submitted by the CPU-C. CPU-M fetches the graph from memory (HBM), unrolls it into WQEs and evaluates dependencies for scheduling. Once ready for scheduling, it tags its own ID on the WQEs and submits them to the NIC Interface or near memory compute. The CPU-M also manages shared completion queues where all the CQEs corresponding to the WQEs that it issued are routed back to. This allows the ME to poll fixed queues and bypass managing individual NICs and multiple completion queues corresponding to different QPs.

\paragraph{NIC Interface} The second component (in blue) is the NIC interface which offloads management of the NIC WQE submission process. It exposes a single FIFO for the CPU-M to post the WQEs and routes the WQE to the Express Doorbell of the correct NIC. Any ME (and NIC interface) can access any NIC (see more details on NIC configurations in Section~\ref{subsubsec:network_chiplet}). 

\paragraph{Near Memory Compute (NMC)} The final component (in yellow) is the near memory compute or reduction engine. This is used to offload data copy and simple sum reduction operations on buffers without involving the Processing Element (PE) grid (Given buffers $S$, $A$, and $B$, NMC can perform operations like $S=A+B$, $A=A+B$, $B=A+B$, $A=B$). Each NMC sustains 128 B/cycle per input, dropping to 96 B/cycle if all NMCs are active. This totals 2.8 TB/s -- more than doubles the aggregate I/O bandwidth -- enabling it to sustain line rate during reduction-based collectives (e.g., AllReduce or ReduceScatter). NMC supports three data types: \texttt{BF16}, \texttt{FP16}, and \texttt{FP32}. The NMC natively does arithmetic in \texttt{FP32} and re-quantizes the data if the datatype differs.

\subsubsection{Network Chiplet}\label{subsubsec:network_chiplet} The network chiplet contains all of the backend off-device I/O capabilities and is integrated directly to avoid PCIe traffic when reading or writing data to caches or HBM.
\athena contains two network chiplets and each chiplet has six custom 800~Gb/s RDMA NICs.
Here we describe some details of the NICs.

\paragraph{Express Doorbells} The classic way of initiating RDMA communication by a NIC is to place work queue entries (WQEs) in queues in host or device memory and hit a doorbell to trigger processing.
The NIC would then perform separate reads - one to pull the WQE itself and others to pull the data from memory.
While this is performance optimal for batched or bulk traffic, it comes with a startup computational cost and latency.
Express doorbells reduce this overhead by treating the WQE write (from NIC interface) itself as the doorbell. In this model, the WQE (with additional metadata like queue-pair ID) is written directly to a special doorbell address and the NIC manages the queuing in its internal memory. This eliminates the additional cost of reading the WQE from an external queue.

\paragraph{QP and WQE limits}
Since WQEs are directly written into the NIC, they need to be stored in NIC memory.
To accommodate this, NICs have a fixed amount of reserved memory to store the WQEs.
This also implies we do not support QP caching (moving less used QPs into HBM) as it would require more complexity and chip area.
However, we have found that for our workloads, these restrictions are well below our typical requirements.

We use custom bits in memory read/write requests to enable features such as separate cache partitions in the compute chiplet for different memory uses. With this, we are able to partition the cache to prevent temporary buffers used by the communication library from polluting the co-resident application data.

\paragraph{Scale-up and Scale-out Communication}
With the pace of change in our workloads, flexibility in rack design was a high priority.
Therefore, we designed this system to adapt to changing network configurations with fungible bandwidth.
Each of the NICs can be wired up to either the scale-up or scale-out switches to reconfigure the system to match our needs.
The initial configuration is to use 2 NICs for scale-out and 8 NICs for scale-up per \athena, all running at 800 Gb/s link speed for a total aggregate bandwidth of 1 TB/s. 
We have an additional 2 NICs available which gives us the option to later increase NIC I/O up to 1.2 TB/s if needed, depending on a balance between application needs, power, and computational cost requirements.



\section{Communication Software Architecture}
\label{sec:comms_architecture}

\begin{figure}[t]
    \centering
    \includegraphics[width=1\linewidth]{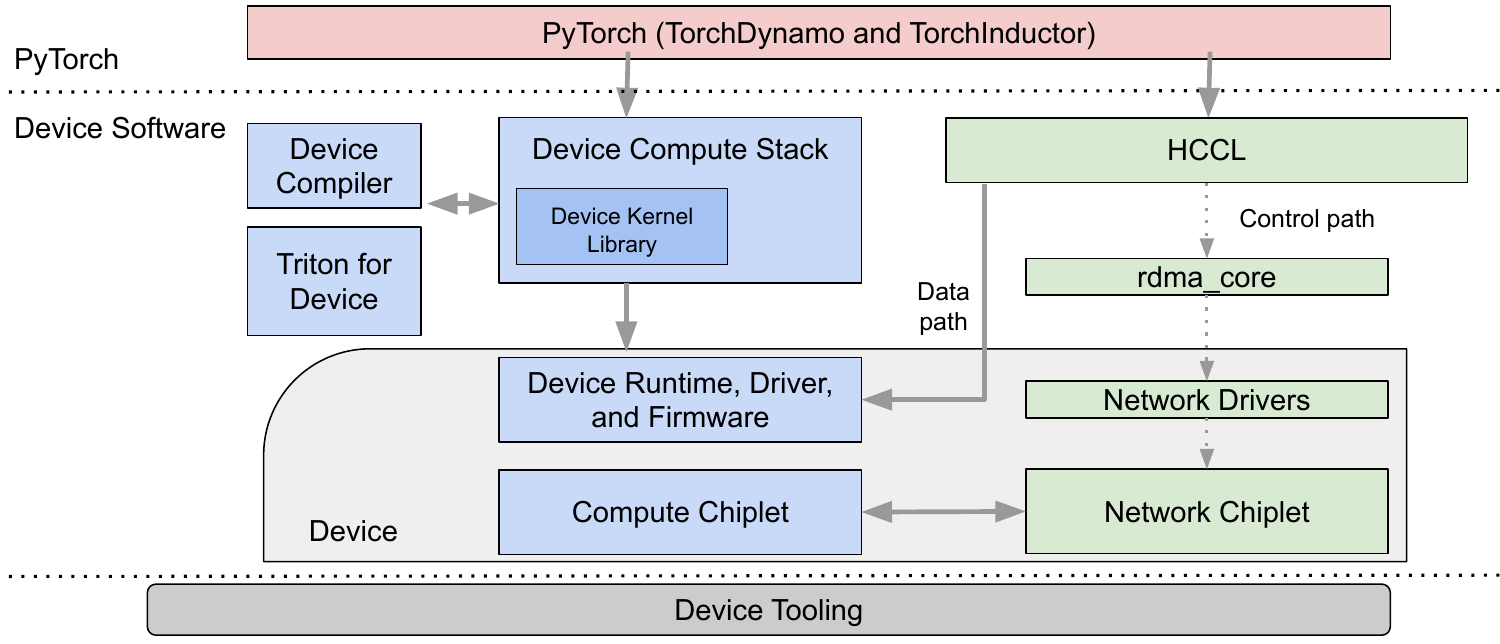}
    \caption{\athena software stack.}
    \label{fig:trainer-stack}
\end{figure}

The software stack for \athena has been closely co-designed with the hardware described in Section~\ref{sec:hardware_architecture} and the needs of our applications. A full view of the software stack can be seen in Figure~\ref{fig:trainer-stack}. 

\subsection{PyTorch}
\label{subsec:pytorch}

PyTorch is the primary interface to the chip both for compute and communication operations. PyTorch Distributed (c10d)~\cite{pytorch-distributed} has been the default communications interface for PyTorch, but recently torchcomms was introduced to better map communication APIs to hardware capabilities and collective communication library (CCL) interfaces. Full details about torchcomms are available in~\cite{torchcomm}. We have added \redact{MTIA}-specific backends to both c10d and torchcomms to enable communication for \athena.

\subsection{Hoot Collective Communications Library (HCCL)}
\label{subsec:hccl}

HCCL is the optimized communication software that translates the communication semantics from PyTorch into operations that are executed in the hardware. As with other communication libraries, it is largely divided into two parts: the control path, which manages/enables the communication through RDMA verbs semantics, and the data path, which represents individual communication operations that are later translated into hardware-specific instructions and thereby generates data traffic in the fabric.

\subsubsection{Control Path}
\label{subsubsec:hccl_control_path}

The HCCL control path manages all of the network resources that are used by the data path later for communication. 
It uses standard RDMA verbs interfaces to manage these resources (e.g., \texttt{ibv\_create\_qp} and \texttt{ibv\_reg\_mr}). 
Each of these resources is created on-demand as the application requires them. Before constructing a collective (see Section~\ref{subsubsec:hccl_data_path} for more details), HCCL needs to determine which QP resources will be needed. 
The control path re-uses existing idle QPs and allocates incremental QPs to fulfill the request, and then provides their descriptors to the data path.
Dynamic QP allocation is a well-known optimization (other collective communication libraries employ a similar control-path design) -- but in HCCL it serves an additional purpose: the express doorbell design described in Section~\ref{subsubsec:network_chiplet} limits the total number of queue pairs that the system can support, making resource reuse important.
While this can theoretically require HCCL to harvest existing QPs to allow them to be reallocated for other purposes, in practice we have not seen a need for this as we can support enough QPs without requiring larger communicator sizes.
Instead we focus on keeping the number of QPs required for a collective algorithm to a minimum and using optimizations other than QP scaling, which can be common in other collective libraries, but do not provide significant benefit in HCCL.

\subsubsection{Data Path}
\label{subsubsec:hccl_data_path}

The HCCL data path covers the components that are responsible for generating and executing the data movement instructions themselves. 
HCCL differentiates itself from traditional communication libraries such as many MPI~\cite{mpi50} implementations as well as other collective communication libraries by offloading not only the data transfer on the NIC itself, but also the execution and dependency processing of the entire communication operation.

\begin{figure}[t]
    \centering
    \includegraphics[width=1\linewidth]{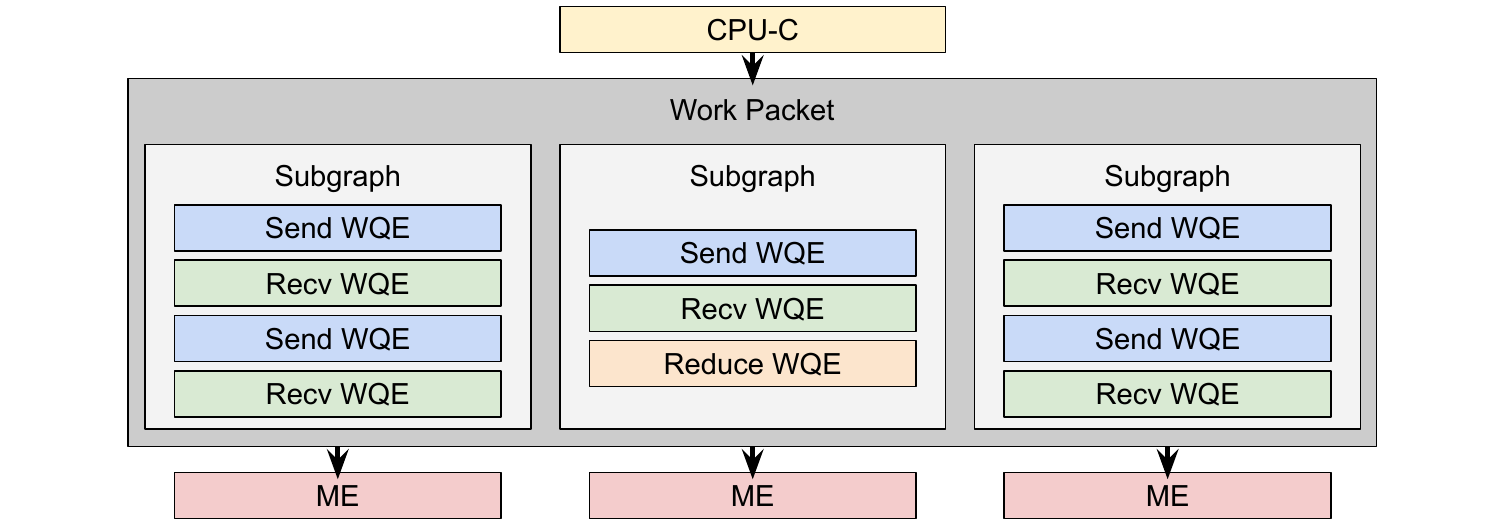}
    \caption{An example of mapping work from CPU-C to CPU-Ms.}
    \label{fig:subgraph-processing}
\end{figure}

The same streaming semantics are used both for communications and compute allowing both to be inserted into the same stream (or they could use cross-stream dependencies if desired). 
A graphical representation of the construction of communication instructions (work packets, subgraphs, and WQEs) is in Figure~\ref{fig:subgraph-processing}.
Streams operate at the granularity of work packets (the equivalent of a compute kernel) which are executed sequentially within a stream. 
Each work packet can be broken down into multiple subgraphs which are parallelized across available MEs (see Section~\ref{subsubsec:me}) to maximize throughput. 
Subgraphs contain a list of work queue elements (WQEs) which describe individual operations to be executed sequentially and include metadata about dependencies between each of the WQEs. 

There are a number of WQE types, but each of them can be put into one of three categories:

\begin{description}
\item[RDMA Transfers] These include \textit{Send}, \textit{Receive}, \textit{Write}, \textit{WriteWithImmediate}, etc. and describe data movement through the RDMA NIC. They contain all of the information necessary to construct an RDMA work request (e.g., QP number, lkey (local key), rkey (remote key), addresses, etc.).
\item[Compute operations] These describe offloaded compute operations that occur in the NMC (see Section~\ref{subsubsec:me}) which can include sums and copies. The copy operation can also function as a copy engine for the communication path.
\item[Subgraph coordination] These describe operations to wait for a condition on a memory location (e.g. the value of a short larger than four) or update / set a value in memory. For instance, two subgraphs can synchronize with each other by using two memory location as a signal pair (using monotonic increasing values) to indicate when the other subgraph can continue to next phase.
\end{description}

\begin{figure}[t]
    \centering
    \includegraphics[width=1\linewidth]{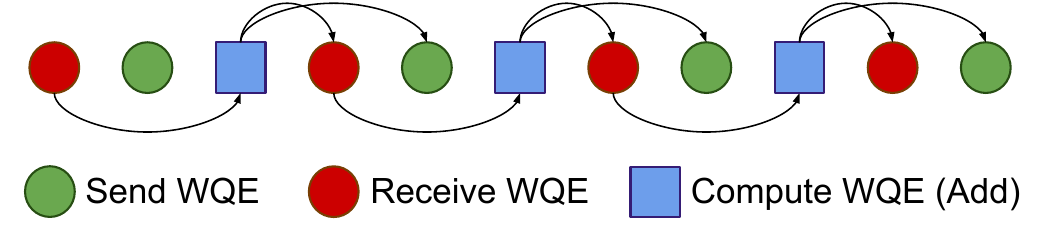}
    \caption{An example of an AllReduce ring algorithm with 4 nodes.}
    \label{fig:allreduce-ring-subgraph}
\end{figure}

Figure~\ref{fig:allreduce-ring-subgraph} depicts a sample subgraph that uses these WQEs. This shows how a series of send, receive, and compute operations can be used to construct a simple ring algorithm for AllReduce~\cite{ar-ring}. 
This also demonstrates how WQEs can express dependencies on other WQEs.
There are five types of dependencies:

\begin{description}
    \item[Fence] Do not start the next WQE until the current WQE has completed.
    \item[Sync] Wait for all previous WQEs to complete before starting the current WQE. This introduces a barrier in the subgraph execution.
    \item[WQE Sync] Wait for a specific WQE to complete before starting the current WQE. Typically used for checking for valid input (pointing at Recv WQE) or buffer release (pointing at Send WQE).
    \item[Receive Sync] Wait for all outstanding receive WQEs to complete before starting the current WQE (i.e., all data for next pass is available).
    \item[Send Sync] Wait for all outstanding send WQEs to complete before starting the current WQE - aka all (temporal) buffers from previous pass are released and available for reuse.
\end{description}

To facilitate multiple dependencies for scheduling a WQE (e.g., a tree based NMC reduction) we have also introduced a NOOP WQE containing only a dependency.

Using send-receive semantics, if there is no matching receive posted when data is incoming to the remote NIC, it will trigger an RNR (receiver not ready) NAK back to the sender, forcing a back-off (wait) with progressively longer retry times.
The retry time is programmable, so there is a compromise between retry time and global/application time-out defined by the max skew between processes. 

To mitigate this, HCCL has a special mode (SendSync) which relies on responder-driven signaling. The requester waits on the signal (using a WAIT WQE) before posting the SEND WQE. 

A further optimization of this is to use the SendSync protocol only on the first transfer in the collective, and let the algorithm guarantee that sufficient receives are posted for any execution scenario, to guarantee that no send will become unmatched by a receive. 
An example of this would be a bidirectional ring with number of receives exceeding pipeline depth.

\section{Optimizations}
\label{sec:optimizations}

Once we achieved a functional \hccl implementation, we spent significant time optimizing the collective library for various application use cases. 
We can generalize most types of optimizations into two categories based on workloads.
While \athena is primarily focused on training for ranking and recommendation models, there is also work done to improve inference performance, both to enable \athena as a chip to support inference itself, and to prepare for future \redact{MTIA} chips, which will be more focused on inference~\anoncite{mtia-blog}.
Details about the performance of these optimizations can be found in Section~\ref{sec:performance}.

\subsection{Training}

Training workloads tend to scale to larger domain sizes and communication is usually bandwidth-bound. These types of workloads were what \athena was originally designed for.
Because of this, the majority of the optimization work to support training workloads is part of the initial work discussed in Section~\ref{subsec:hccl}.
However, one additional area to point out is parallel ME Processing.

\athena has 16 MEs to allow collectives to be broken down into multiple subgraphs that can execute in parallel while maintaining the offloading capabilities provided by having a separate ME engine.
Because each ME is a single-core, we use multiple MEs in order to most efficiently distribute work to each of the NICs and NMCs.
\hccl is able to use all of these MEs for a variety of potential optimizations for the collective algorithm implementations.
For example, it is common for an AllReduce collective using rings to actually be broken into multiple rings with data sharded across them.
The same can be true even for simpler collective implementations such as AllToAllv to both maximize injection rate and/or spread chunks of data across multiple subgraphs.

\subsection{Inference}
\label{subsec:inference}

Inference workloads often require prioritizing a different set of communication use cases. 
Specifically, their workloads are often smaller node counts and smaller data sizes, meaning they are more latency sensitive.
While these workloads are not the primary use case for \athena, typically there is interest in optimizing for inference use cases and in this section, we describe some of the work done toward this optimization.

\subsubsection{One-sided Communication}
\label{subsubsec:one-sided}

One-sided communication APIs have been available in many forms including remote memory access APIs in MPI-2~\cite{mpi20} and again in MPI-3~\cite{mpi30}, various SHMEM~\cite{barriuso1994shmem,openshmem2020spec,nvshmem} libraries, and others. 
Some of the relevant APIs available to us through these interfaces are \texttt{put} (issue an RDMA \texttt{write} directly) and synchronization semantics such as \texttt{flush} or \texttt{fence} to ensure local completion and ordering as well as a way of signaling a remote process that data has landed (e.g., \texttt{signal} and \texttt{wait}).
These interfaces are often used in fused kernels that combine compute and collective operations in a single kernel that runs on the PE grid.
This minimizes latency by eliminating the computational cost of launching additional kernels.

For \athena, implementing one-sided communication presents some design and performance tradeoffs.
As previously described, communication flows through the MEs, which are responsible for executing subgraphs and dispatching work to the NICs and NMCs.
For one-sided communication, each communication round is a single NIC WQE, so this model no longer makes sense due to the computational cost of issuing each as a subgraph.
On the other hand, because of the shared completion queue described in Section~\ref{subsubsec:me}, it is difficult to efficiently monitor completion of individual WQEs on the PE grid.
To manage this, we use a hybrid model where the PE submits WQEs directly through the express doorbell, but the ME receives the completion notification and manages a counter of the completions that the PE can read to know when its messages have been delivered.
The host is responsible for setting up the communication control path (i.e., queue pairs and memory regions) and handing those over to the device, which is managed through \hccl APIs that can have an adaptable implementation from one hardware generation to the next as capabilities and requirements shift.
An example of this workflow is found in Figure~\ref{fig:pe-triggered-example} under the column "One-Sided".

\subsubsection{Device-resident Collectives}
\label{subsubsec:device_resident_collectives}

A common pattern of compute and communication in inference workloads for mixture of experts (MoE) kernels is to have a routing kernel that determines which expert(s) to use, followed by an \texttt{AllToAllv} collective whose metadata (i.e., sizes and offsets) depend on the output of the routing kernel.
Traditionally each of these would be submitted and executed eagerly, either in the same stream to express the data dependency or across two streams with an explicit cross stream dependency.
However, this adds significant latency because HCCL cannot construct the subgraphs for the \texttt{AllToAllv} collective until the previous kernel is complete.
Then the collective is submitted, copied, and scheduled on the device, all of which would be exposed as they are no longer overlapped with the compute kernel that precedes them.
To address this, a variant collective was proposed called \texttt{AllToAllvDynamic}~\cite{100kgpus} which, in addition the existing parameters of \texttt{AllToAllv}, takes pointers to device memory where the final metadata will be written.
This allows the collective library to modify the WQEs before they are sent out without requiring the entire host-side software stack to be involved.
Figure~\ref{fig:pe-triggered-example} demonstrates this workflow in the column labeled "Device-Resident".

\subsubsection{Device-triggered Collectives}
\label{subsubsec:device_triggered_collectives}

\texttt{AllToAllvDynamic} removes the computational overhead of generating the collective and copying it to the device, however, it still leaves the scheduling latency exposed.
For compute kernels, this is typically hidden by fusing together multiple compute kernels in graph mode allowing them to execute as one large kernel which avoids the scheduling latency that would normally occur between each.
However, though compute kernels and collective operations can occur in the same stream, they cannot directly be fused because they are not executed by the same hardware.

We explored another concept called device-triggered collectives where the caller can attach a memory address and comparator to a collective to indicate that the collective should not begin until the memory address fulfills the requirement of the comparator (e.g., a memory address becomes non-zero).
Additionally, the user can provide a second address and value to be set when the collective is done.
With these tools, compute kernels can be compiled in graph mode and still incorporate accelerated communication on the MEs by scheduling the collective on another stream where it is immediately executed, but then blocked until the compute kernel signals that it is ready.
In the subgraph, this is implemented by adding additional WQEs where necessary to perform the indicated comparator and wait for it to be true.
In the case of multiple subgraphs, these WQEs will need to be added to each of the subgraphs that are immediately scheduled.

\begin{figure}[t]
    \centering
    \includegraphics[width=1\linewidth]{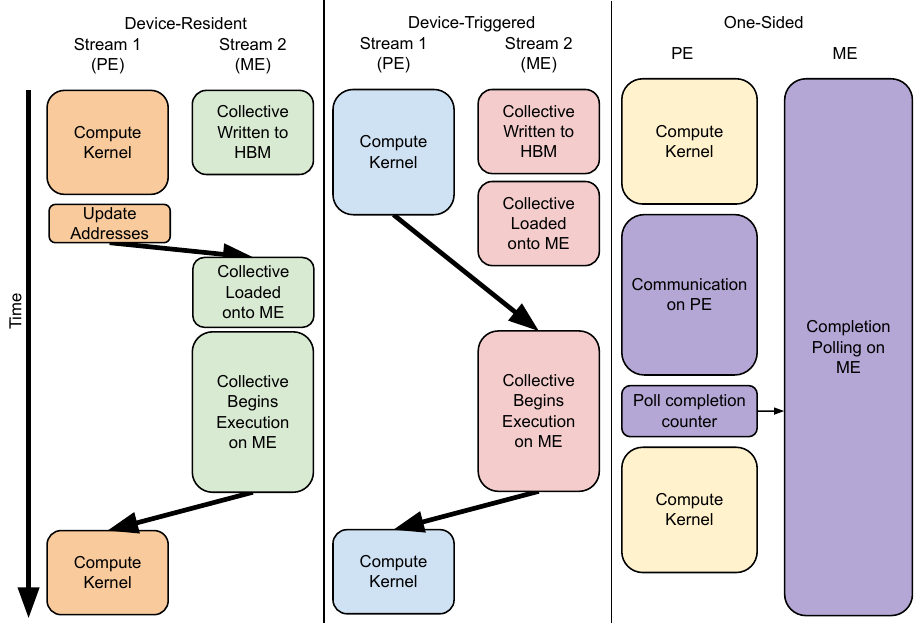}
    \caption{An example of a PE triggered execution flow.}
    \label{fig:pe-triggered-example}
\end{figure}

Upon completion, the CPU-M needs to indicate back to the PE grid that the next compute work is unblocked.
This can either be done in an existing subgraph via a WQE to set the corresponding value or via a dedicated subgraph if there are multiple subgraphs that need to avoid prematurely writing to a shared value.
In the case of multiple subgraphs, HCCL uses internal semaphores to ensure ordering is preserved. 
Figure~\ref{fig:pe-triggered-example} visually compares this method (labeled "Device-Triggered") to the other two methods described here.

The primary benefit of this optimization is to maintain a single compute kernel when executing in graph mode.
Graph submission and launching is a non-negligible computational overhead that can be avoided if the compute and communication work can be executed in a single kernel launch.
By embedding the collective semaphore into the compute kernel and pre-launching the collective in a separate stream, we can avoid graph breaks that would force multiple CPU-C scheduling computational overheads.

Scheduling the collective on a parallel stream also enables the MEs to pre-post parts of the work until the compute has signaled the ME to start collective execution.
Upon scheduling, the ME can also pre-post receive WQEs before the wait WQEs mentioned above since these do not incur in any data exchange until a matching send is posted on the remote node.
This enables a two-fold optimization: 1) we save execution time in the ME since fewer WQEs need to be posted in the exposed time of the collective and 2) we mitigate the occurrence of "Receiver not Ready" NAKs since, for a sufficiently large time window between scheduling and signaling, we can expect all ranks to have pre-posted their receive WQEs.
\section{Performance}
\label{sec:performance}


To better understand HCCL performance on \athena, we use synthetic benchmarks as well as end-to-end runs.
HCCL performance can be observed from both the host-side and device-side, each with a different impact on workload behavior.

These experiments were all run using the number of \athena accelerators specified along with an AMD EPYC 9334 32-core processor as the host. Other specifics of the hardware setup can be found in Section~\ref{sec:hardware_architecture}.

\subsection{Workload analysis}

\begin{figure}[t]
    \centering
    \includegraphics[width=1\linewidth]{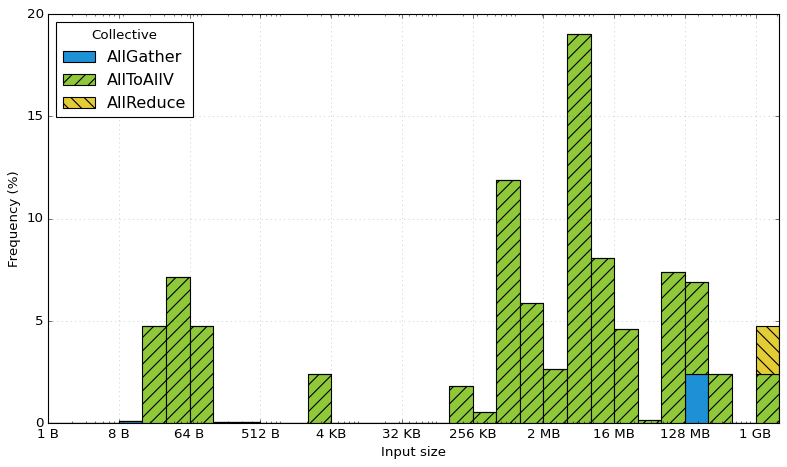}
    \caption{Collective distribution for a ranking and recommendation workload at 40-ranks scale}
    \label{fig:mai-coll-dist}
\end{figure}

\begin{figure}[t]
    \centering
    \includegraphics[width=1\linewidth]{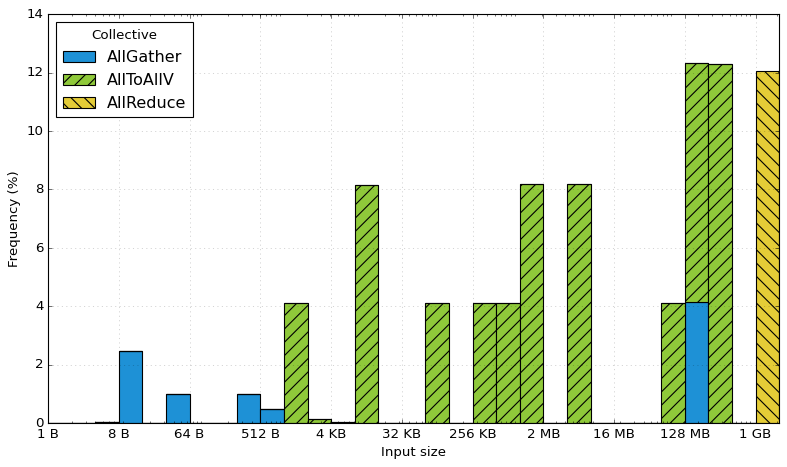}
    \caption{Collective distribution for a ranking and recommendation workload at 256-ranks scale}
    \label{fig:e2v-coll-dist}
\end{figure}

Given \athena's focus on ranking and recommendation, we perform an analysis from real training workloads.
We show the distribution of collective data sizes from two different workloads in Figures~\ref{fig:mai-coll-dist} and~\ref{fig:e2v-coll-dist}.
Each model is running its normal configuration, including hyperparameters and job size.

We observe a strong tendency in these models preferring medium to large collectives for execution, specifically large AllToAllv (for embedding exchange in each batch), AllReduce (for gradient accumulation on backwards pass) and AllGather (for the loss optimizer).
Therefore, we focus on these collectives as our optimization goal and benchmarking metric during \hccl's development.

\subsection{Host-side Performance}

Due to our compiled communications approach, the host is required to build all communication graphs before copying them to the device.
In practice, host-side generation time is hidden from the critical path due to the pipelining effect on device-side execution, and this does not pose any performance limitations on our target training workloads.

\begin{figure}[t]
    \centering
    \includegraphics[width=1\linewidth]{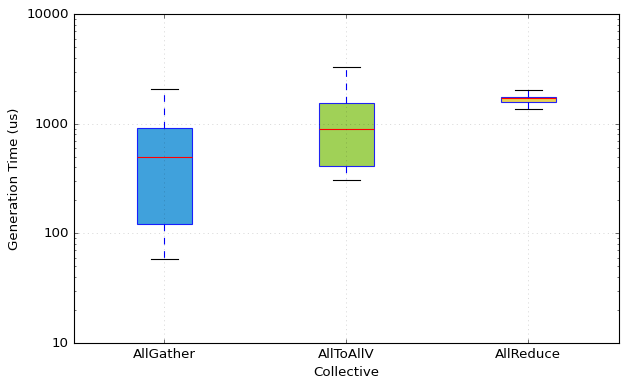}
    \caption{Host-generation time for a ranking and recommendation workload at 256-ranks scale}
    \label{fig:e2v-gen-time}
\end{figure}

Host-side generation is composed of graph generation, device-memory setup, as well as any required cross-stream dependency injection and collective kernel enqueuing.
We capture runtime metadata from target workload runs at different scales to observe the execution time of each collective.
Figure~\ref{fig:e2v-gen-time} presents a box-plot with the observed distribution of generation times during full execution of the training job.
Some collectives see varied generation times, which is related to different input tensor sizes being used and therefore selecting different collective algorithms.


We do not observe any exposed host-time execution from this generation when validating these workloads, but more latency-sensitive workloads are subject to host computational overhead.
To mitigate this effect, we also employ a work packet caching strategy where previously compiled collectives can be reused (that is, device-memory is not released upon the collective completion).
This effectively removes the graph generation time and is limited to re-enqueuing the collective back on the stream for execution.
We observe less than $10~\mu\mathrm{s}$ for such operations, which is static and does not depend on any of the execution parameters (such as job size, input size).

\subsection{Control-core Computational Overhead}

\begin{figure}[t]
    \centering
    \includegraphics[width=1\linewidth]{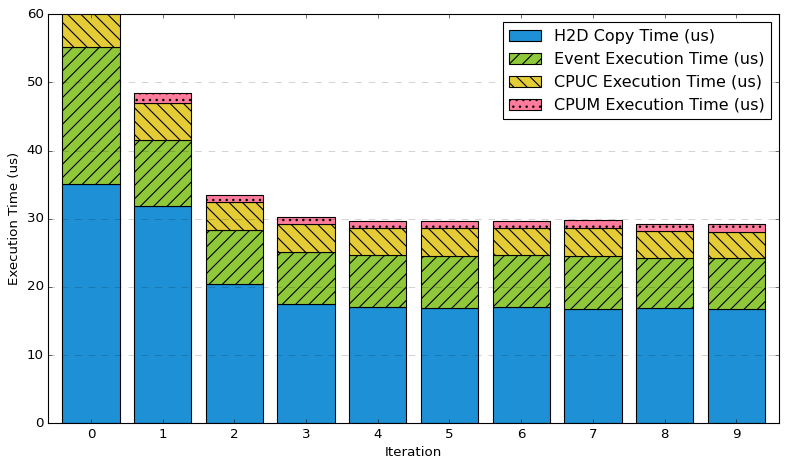}
    \caption{Breakdown of execution time per component of execution}
    \label{fig:ccp-breakdown}
\end{figure}

\athena uses a control core (CPU-C) as the main entry-point for device operations. 
This behavior can be observed as important to collective performance, because CPU-C scheduling time can dictate when a CPU-M initiates communication, as displayed in Figure~\ref{fig:subgraph-processing}.
We designed a synthetic benchmark which enqueues a single work packet with a single subgraph which contains a single reduction WQE to copy 4KB per iteration.
Leveraging this type of subgraph enables minimizing latencies and variation from other factors, such as network transfers.
We display a total of 10 iterations in Figure~\ref{fig:ccp-breakdown}, with the initial 3 iterations demonstrating non-pipelined performance, while iteration 4 and beyond represents more common scenarios.
The reduction WQE runs an entire-pass of the CPU-M state machine while keeping execution time to a minimum, allowing us to mimic a very low-latency collective for measurement purposes.

As typical in GPU systems, \textit{Stream Events} are also available on \athena, which are used by models to perform cross-stream dependency validation, as well as used for collecting asynchronous execution timings.
Given their prevalence across workloads and their usage in our end-to-end benchmarks, we use them as the highest level of granularity for this benchmark.
Counters are also available on each work packet and subgraph which can provide indication of when a specific component has observed a work packet's execution.
With this information and the component relationship of $T(CPU M) \subseteq T(CPU C) \subseteq T(Event)$, we compute the individual computational overhead of each component in Figure~\ref{fig:ccp-breakdown}.
Once the event pipeline is full, we observe a static computational overhead of $17.1 \pm 0.3 \mu \mathrm{s}$ for the Host-to-Device (H2D) copy time, $3.6 \pm 0.1 \mu \mathrm{s}$ for event handling and $2.9 \pm 0.1 \mu \mathrm{s}$ for CPU-C work dispatching, while the CPU-M execution time was the smallest at $1.1 \pm 0.1 \mu \mathrm{s}$.
Since CPU-M execution is what actually represents collective execution, we can conclude that Event timers will incur non-negligible computational overhead close to $6 \mu \mathrm{s}$ in total during workloads and end-to-end benchmarking.
H2D copies, while slow, are hidden in practice by copying on a parallel stream while some other work is executing (using the device copy engines), which makes it negligible during real workloads.

\subsection{End-to-end Performance and Scalability}


We now quantify the end-to-end performance of \hccl on \athena, testing different collectives as exposed via \hccl's API and sweeping across different job and input sizes.
Given the computational overheads presented in Figure~\ref{fig:ccp-breakdown}, we collect our data from fine-grained event timers inserted before and after each collective's execution, bypassing all metadata copying to the device and any scheduling computational overhead from the control core.
We bypass these parts because they would typically be overlapped with previous compute when executing as part of a model.
Collectives are enqueued back-to-back, using 100 warm-up iterations and 1000 benchmark iterations.

We employ rack-level allocation to ensure proper system load-balancing and node distribution.
Job sizes smaller than the rack size (16) are guaranteed to be scale-up only measurements, while jobs larger than rack size are guaranteed to have contiguous rank distribution on each rack, and all racks are full.
A job with 64 ranks is then guaranteed to use 4 racks, with ranks 0-15 in rack 0, ranks 16-31 in rack 1, and so on.

\begin{figure}[t]
    \centering
    \includegraphics[width=1\linewidth]{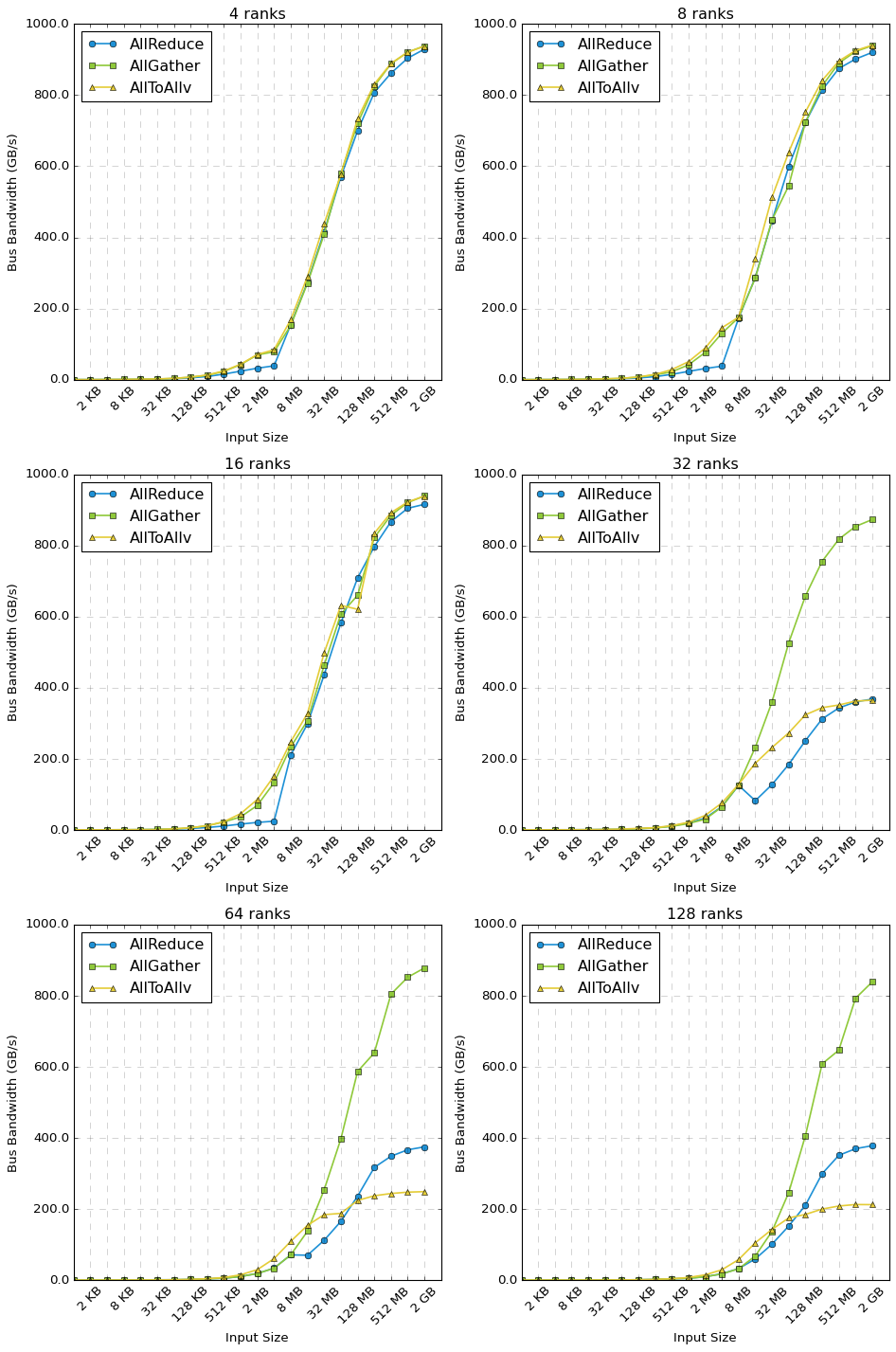}
    \caption{End-to-end test sweep across different job sizes}
    \label{fig:hccl-test-sweep}
\end{figure}

Given the workloads observed in Figure~\ref{fig:mai-coll-dist} and Figure~\ref{fig:e2v-coll-dist}, we focus our testing on AllToAllv, AllGather, and AllReduce.
Figure~\ref{fig:hccl-test-sweep} presents our benchmark sweep over these collectives up to 128 ranks, showcasing the achieved on-the-wire bandwidth of each collective.
 
Overall, we see good performance from \hccl regarding full system bandwidth usage.
With a total theoretical bandwidth of 1~TB/s (800~GB/s scale-up, 200~GB/s scale-out), we observe up to 940 GB/s in collective performance for collectives that remain within a single scale-up domain.
As the collective size grows larger than a rack, the total bandwidth bends closer to the maximum scale-out bandwidth based on the percentage of data that must flow out of a rack (i.e., 50\% for AllToAllv for 32 ranks).
The dedicated offloaded collective path via the ME offers decoupled HW support for collective execution and is thus capable of saturating system resources.
Larger job sizes also demonstrate good resource usage despite the fact that most bandwidth is scale-up only.

For larger-scale collectives, we employ topology-aware algorithms to maximize system utility and properly balance scale-up and scale-out usage during collective execution, which results in a lower execution time.
This is especially noticeable for AllGather, which is capable of achieving 838 GB/s at 128-rank scale despite only having 200 GB/s available between scale-out ranks.
We minimize message exchanges on the scale-out network and instead use the over-provisioned scale-up bandwidth to exchange data among ranks in the same rack, which results in higher throughput.

\subsection{Compute Communication Overlap}

\begin{figure}[t]
    \centering
    \includegraphics[width=1\linewidth]{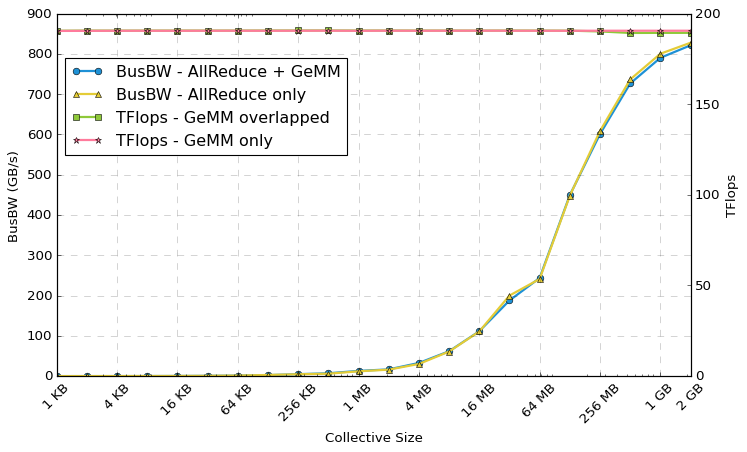}
    \caption{Variance of GEMM performance when scaling collective size at 16-ranks scale.}
    \label{fig:compute-comms-overlap}
\end{figure}

Computation and communication overlap is a common scenario used by models.
We use the PARAM~\cite{param} benchmark suite to validate the performance of \athena on such scenarios, and the results can be found in Figure~\ref{fig:compute-comms-overlap}.
This benchmark measures the throughput of a large GEMM which saturates the compute grid while running 100 back-to-back collectives of the same size on a parallel stream.
Similar to Figure~\ref{fig:hccl-test-sweep}, we sweep all relevant collectives among different data sizes, but keep the execution at 16-ranks scale.
We observe minimal performance degradation on GEMM throughput, with a variance of up to 1 TFlop (around $0.5\%$), while collective performance grows similar to the measurements without any background computation.
We estimate the drop in compute performance to be associated with higher HBM usage from the NICs concurrently accessing memory.

The sustained compute bandwidth demonstrates the relevance of our offloaded communication path which frees up the compute cores from performing network operations.
For workloads that overlap collective and compute operations at large tensor sizes, there is no requirement to load-balance resource utilization by each kernel.

\subsection{Inference Performance Optimizations}

We focus now on analyzing the optimizations mentioned in Section~\ref{subsec:inference} for inference workloads.

\subsubsection{Uni-Directional Communication Message Latency}

\begin{figure}[t]
    \centering
    \includegraphics[width=1\linewidth]{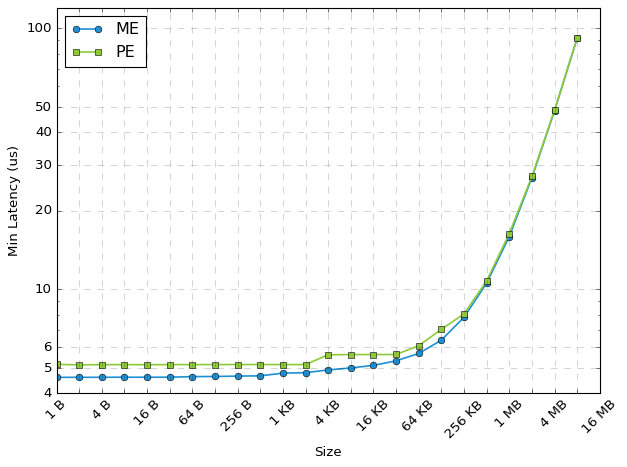}
    \caption{RDMA Write WQE RTT comparison between PE vs ME}
    \label{fig:inference-put-flush-bench}
\end{figure}

We design an end-to-end benchmark to compare the computational overhead of issuing WQEs from the PE vs ME, shown in Figure~\ref{fig:inference-put-flush-bench}.
We observe a static computational overhead of around $450~\mathrm{ns}$ when submitting WQEs directly from the PE grid, which is due to the computational overhead of building WQEs on the device, bookkeeping for operation correctness, and obtaining completions routed to the PE through an ME. This demonstrates the efficacy of submitting work from the PE grid to enable fused kernels.

\subsubsection{Device-Triggered Collectives}
\label{subsubsec:device_collectives}

\begin{figure}[t]
    \centering
    \includegraphics[width=1\linewidth]{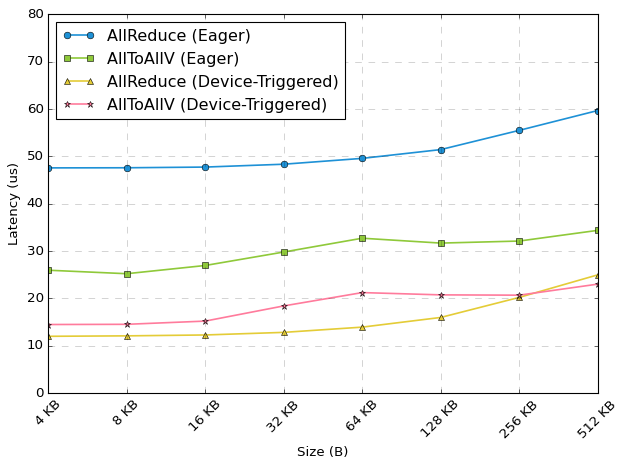}
    \caption{Device-triggered collectives performance compared to eager execution}
    \label{fig:inference-pe-triggered}
\end{figure}

\begin{figure}[t]
    \centering
    \includegraphics[width=1\linewidth]{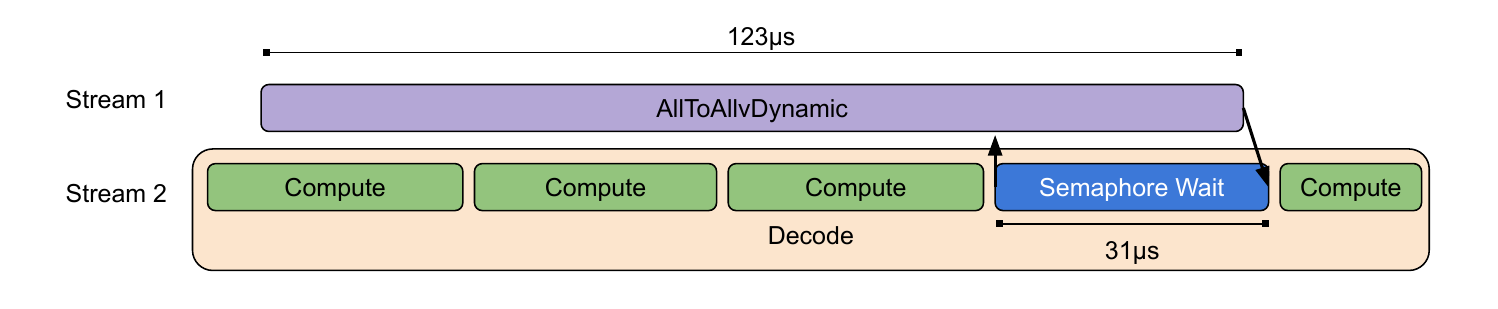}
    \caption{Execution example of Device-triggered execution.}
    \label{fig:pe-triggered-kineto-trace}
\end{figure}

Sections~\ref{subsubsec:device_resident_collectives} and~\ref{subsubsec:device_triggered_collectives} describe optimizations that allow collectives to be executed in fused kernels with compute operations in graph mode.
We measured the performance benefit of these two optimizations together as they are less impactful separately.
In Figure~\ref{fig:pe-triggered-kineto-trace}, we can see an \texttt{AllToAllvDynamic} collective that is scheduled simultaneously with a series of fused kernels.
The collective itself appears to execute for $123 \mu \mathrm{s}$ on the trace, but in the compute stream, it is actually only executing for $31 \mu \mathrm{s}$ as most of the execution time of the collective is actually blocked while waiting for the compute kernel to signal readiness.
This avoids all of the computational overheads demonstrated in Figure~\ref{fig:ccp-breakdown} that would be injected by having to schedule and execute the collective after the compute kernel is finished.
Figure~\ref{fig:inference-pe-triggered} shows a performance comparison of device-triggered collectives against their eagerly executed counterparts showing significant latency improvement from avoiding the compute necessary for setup.

\subsubsection{PE Collective Kernels}
\begin{figure}[t]
    \centering
    \includegraphics[width=1\linewidth]{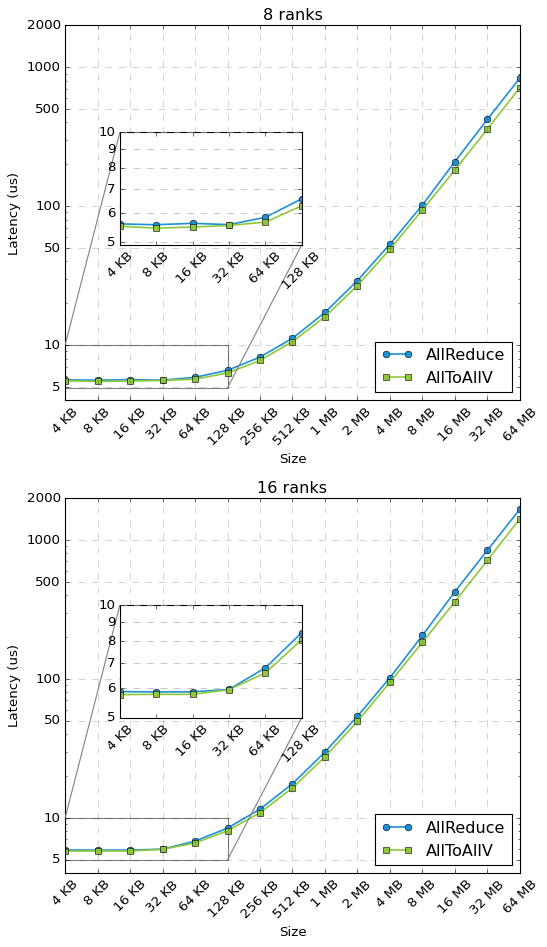}
    \caption{Performance of PUT-based collective implementations}
    \label{fig:inference-put-bench}
\end{figure}

We implement AllReduce and AllToAllv kernels on the PE grid to be used as a direct replacement for HCCL offloaded collectives.
We take this approach based on the numbers presented in Figure~\ref{fig:ccp-breakdown}, which indicate some computational overhead from scheduling to the ME itself.
Figure~\ref{fig:inference-put-bench} shows the performance of these kernels when running end-to-end and at the job sizes within the scale-up domain for running inference workloads.
We achieve sub-$6 \mu \mathrm{s}$ collective execution for both AllToAllv and AllReduce, with AllReduce containing additional synchronization requirements due to the reduction step.

Both collectives are implemented in an ``all-to-all" style, with all ranks writing to all ranks over the network.
This also enables different cores to issue parallel messages automatically without synchronization, which helps mitigate skew and message enqueue computational overhead for such fine-grained workloads.

At this stage, our solution is mostly bound by the RTT time of the network, with little optimization to be done at the SW level.
Further improvements on these numbers would likely rely on updates to the network itself to improve end-to-end message delivery time.
While not mentioned in this work, this functionality also enables fused-kernel design which would reduce the control core computational overhead and further cut execution time during end-to-end inference (along with further optimizations described next).

\section{Related Work}
\label{sec:related}

Collective communication has a long history in HPC through the Message Passing Interface (MPI) standard~\cite{mpi50}, with widely deployed implementations including MPICH~\cite{mpich-site} and Open MPI~\cite{open_mpi}.
These libraries provide rich collective semantics and have been extensively optimized for traditional HPC networks and interconnects.
However, MPI implementations are primarily designed for CPU-based communication and do not natively understand accelerator memory hierarchies or device-side execution models.
As accelerators have become the dominant compute platform for AI workloads, a new generation of collective communication libraries have emerged to address these gaps.

NVIDIA's NCCL~\cite{nccl} was purpose-built for GPU collective communication and has become the de facto standard for GPU-based distributed training.
NCCL uses a kernel-based execution model where collective algorithms run as persistent GPU kernels that directly post work to the NIC through GPUDirect RDMA.
This approach ties collective execution to GPU compute resources: the streaming multiprocessors (SMs) running the NCCL kernel are unavailable for application compute during communication.
NCCL mitigates this by using a small number of SMs, but the coupling between compute and communication resources remains.
In contrast, HCCL's message engines are dedicated hardware that execute collectives without consuming any PE grid resources, enabling compute-communication overlap as shown in Section~\ref{sec:performance}.

\begin{figure}[t]
    \centering
    \includegraphics[width=1\linewidth]{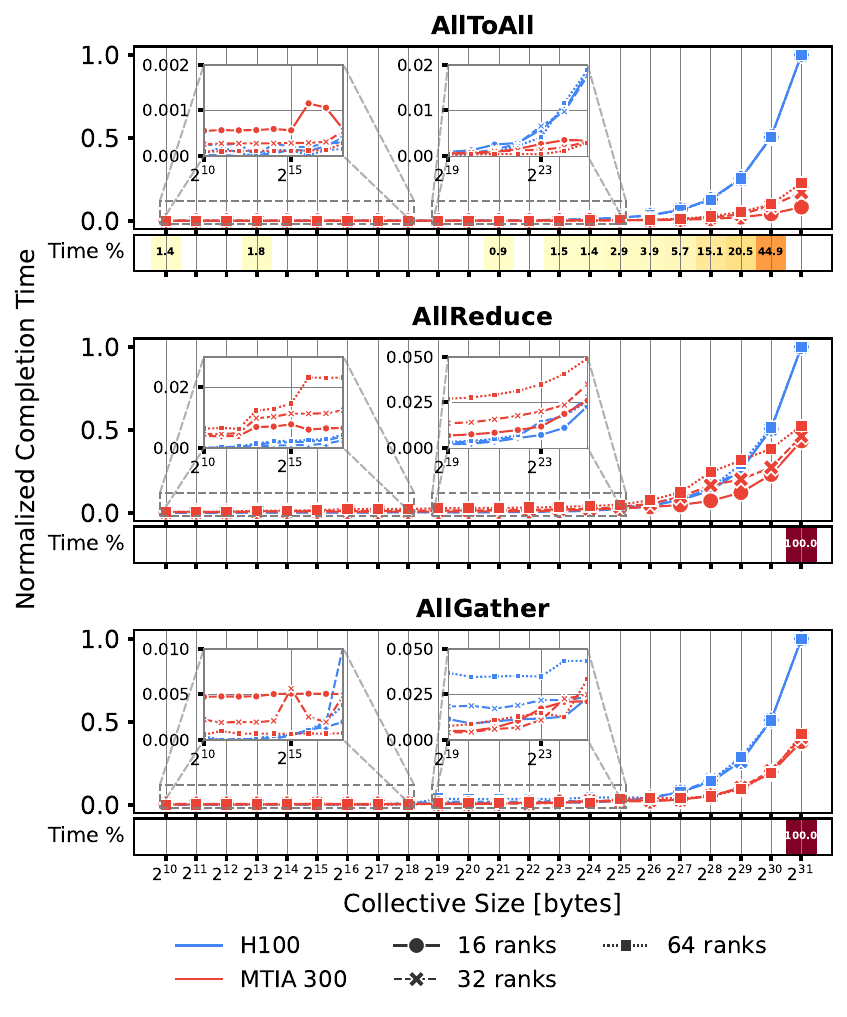}
    \caption{Performance of collective operations. ``Time \%'' represents the ratio of execution time for collectives with different message sizes in our workloads.}
    \label{fig:comm-param-perf}
\end{figure}

We have previous work~\cite{athena_isca} that describes the relative performance of various workloads on MTIA 300 vs NVIDIA H100 GPUs.
Figure~\ref{fig:comm-param-perf} from that paper shows a particularly notable speedup when using 16 or more accelerators or message sizes over 16 MB, which we attribute to MTIA 300's larger scale-up domain size and 2.2× higher scale-up bandwidth. 
For small message sizes, H100 with NCCL currently tends to outperform MTIA 300 with HCCL. 

NVIDIA has also explored offloading collective operations away from GPU SMs through two complementary technologies.
SHARP~\cite{sharp} performs in-network reductions on InfiniBand switches, enabling allreduce and other
collectives to be computed as data traverses the network fabric rather than at the endpoints.
The BlueField DPU~\cite{bluefield} takes a different approach by providing a programmable network processor co-located with the NIC that can execute communication and collective logic independently of the host or GPU.
Both aim to reduce the compute resources consumed by communication, a goal shared by HCCL's message engine design; however, SHARP and BlueField are external to the accelerator itself, whereas HCCL's offload engines are integrated on the chip package and operate on accelerator-local memory without traversing PCIe or host interfaces.

AMD's RCCL~\cite{rccl} provides NCCL-compatible APIs for AMD GPUs running on the ROCm platform.
Like NCCL, it uses a kernel-based execution model, but must also account for the differences in AMD's interconnect hierarchy (i.e., Infinity Fabric vs. NVLink).
Intel's oneCCL~\cite{oneccl} serves a similar role for Intel accelerators including Gaudi, providing a unified API across Intel's hardware portfolio.

Microsoft's MSCCL~\cite{msccl} and its domain-specific language MSCCLang~\cite{mscclang} take a compiler-driven approach to collective algorithm design.
Rather than hand-tuning algorithms for specific topologies, MSCCL allows users to express collective algorithms as programs that are then compiled and lowered onto the target hardware.
This shares a philosophical similarity with HCCL's compiled communication model in that both separate the description of a collective from its execution, though MSCCL targets GPU kernels while HCCL targets dedicated message engine subgraphs.

Gloo~\cite{gloo} is a CPU-focused collective library used in PyTorch's distributed runtime, primarily for parameter server and CPU-based coordination tasks such as distributed barrier and broadcast operations during initialization.

\section{Conclusion}
\label{sec:conclusion}

We have presented HCCL, a collective communication library co-designed with \redact{Meta}'s \athena accelerator to exploit its integrated networking hardware. By offloading collective execution entirely to dedicated message engines and near-memory compute units, HCCL achieves large compute-communication overlap --- a property that is difficult to realize in kernel-based collective libraries where communication consumes device compute resources.

HCCL's compiled communication model, in which the host generates subgraphs that are autonomously executed by the MEs, enables efficient use of \athena's RDMA NIC hardware while keeping the control path flexible enough to support topology-aware algorithm selection across the chip's asymmetric scale-up and scale-out network. Our performance results demonstrate that this approach can saturate available bandwidth for training workloads while also achieving the low latency required for inference through one-sided communication and device-triggered collective designs.

As \redact{Meta}'s \redact{MTIA} chip family continues to evolve toward future generations with greater emphasis on inference workloads~\anoncite{mtia-blog}, the communication primitives and architectural patterns established in HCCL --- particularly the one-sided PE-initiated paths and device-triggered collectives---provide a foundation for continued optimization. The co-design methodology between the communication library and the underlying hardware has helped achieve competitive performance on a first-generation integrated networking architecture, and we expect this approach to continue for future \redact{MTIA} development.

\section{Acknowledgements}

The authors would like to thank the numerous people across the MTIA networking, silicon, communications, compute, and other teams who contributed to the success of this project who could not be listed as authors. A complete list of contributors to the program can be found in~\cite{athena_isca}.

Some of the text for this paper was written using Meta's Muse Spark 1.1 and Claude Opus 4.6.

\bibliographystyle{IEEEtranS}
\bibliography{refs}

\end{document}